\documentclass[english,aps,prl,twocolumn,10pt,showpacs]{revtex4-1}
\usepackage[T1]{fontenc}
\usepackage[latin9]{inputenc}
\usepackage{units}
\usepackage{amssymb}
\usepackage{graphicx}

\makeatletter

\newcommand\PiStrut{\rule[0.8em]{0pt}{0.03em}}

\makeatother

\usepackage{babel}
\begin{document}

\title{Microwave state transfer and adiabatic dynamics of magnetically trapped
polar molecules}

\author{Benjamin K. Stuhl}
\email{stuhl@jila.colorado.edu}
\author{Mark Yeo}

\author{Brian C. Sawyer}
\altaffiliation{Present address: Ion Storage Group, National Institute of Standards and Technology, Boulder, CO 80305, USA}

\author{Matt Hummon}
\author{Jun Ye}
\affiliation{JILA, National Institute of Standards and Technology and University
of Colorado, Department of Physics, 440 UCB, Boulder, Colorado 80309,
USA}


\begin{abstract}
Cold and ultracold polar molecules with nonzero electronic angular
momentum are of great interest for studies in quantum chemistry and
control, investigations of novel quantum systems, and precision measurement.
However, in mixed electric and magnetic fields, these molecules are
generically subject to a large set of avoided crossings among their
Zeeman sublevels; in magnetic traps, these crossings lead to distorted
potentials and trap loss from electric bias fields. We have characterized
these crossings in OH by microwave-transferring trapped OH molecules
from the upper $\left|f;\, M=+\frac{3}{2}\right\rangle $ parity state
to the lower $\left|e;\,+\frac{3}{2}\right\rangle $ state and observing
their trap dynamics under an applied electric bias field. Our observations
are very well described by a simple Landau-Zener model, yielding insight
to the rich spectra and dynamics of polar radicals in mixed external
fields.
\end{abstract}

\maketitle

Cold polar molecule experiments have made a remarkable transformation
over the past decade, progressing all the way from the first demonstrations
of basic motional control of molecules \cite{Weinstein1998,PhysRevLett.83.1558,PhysRevLett.91.243001}
to studies in ultracold chemistry \cite{Ospelkaus12022010}, polar
collisions \cite{Ni2010,2010arXiv1008.5127S,Miranda2011}, and precision
measurement \cite{PhysRevLett.96.143004,*PhysRevA.74.061402,Hudson2011}. Closed-shell molecules
at low temperatures and long ranges present nearly featureless electric
dipoles \cite{PhysRevA.63.052714,PhysRevA.72.032717}, yielding universal
scattering behavior and the prediction of generic dipolar condensates
\cite{PhysRevLett.85.1791} or crystals \cite{PhysRevLett.100.050402}.
At shorter ranges, the quantum statistics and chemical nature of the
specific species comes to the forefront \cite{Ospelkaus12022010,PhysRevLett.101.203203}.
Polar \emph{radicals} behave in even richer fashions: topological
crystals \cite{Micheli2006}, half-integer vortices generated by conical
intersections \cite{PhysRevLett.103.083201}, and spin-dependent chemistry
\cite{PhysRevA.83.032714} have been predicted, and the utility of
using magnetic fields to trap radicals while performing electric-dipole-dependent
studies has already been demonstrated \cite{PhysRevLett.98.253002,PhysRevLett.101.203203,PhysRevLett.105.153201}.
Motivated by these predictions, several groups are pursuing the production
of ultracold radicals such as RbSr \cite{PhysRevLett.105.153201,PhysRevA.82.042508}
and LiYb \cite{PhysRevA.84.022507}.

While the combination of electric and magnetic dipoles enables fascinating
new phenomena, it also comes with substantial complications. In particular,
simple linear Zeeman spectra are converted to tangles of interwoven
avoided crossings by the application of remarkably small transverse
electric fields. In this Letter, we report the experimental observation
of these crossings in the polar radical OH and their dramatic impact
on dynamics in a magnetic trap. In its $^{2}\Pi_{3/2}$ ground electronic
state, stationary OH has a magnetic dipole moment of $2\mu_{B}$ (where
$\mu_{B}$ is the Bohr magneton) and an electric moment of 1.67 D
(0.65 a.~u.); rotational averaging reduces each of these in the laboratory
frame to $\mu_{i}^{(eff)}=\mu_{i}\frac{M\bar{\Omega}}{J\left(J+1\right)}$,
where $J$ is the total angular momentum, $M$ is its lab-fixed projection,
and $\bar{\Omega}$ is the magnitude of $J$'s projection on the internuclear
axis. Each rotational level of OH contains two Zeeman manifolds of
opposite parity, $\left|e;\, M\right\rangle $ and $\left|f;\, M\right\rangle $;
with no applied fields, these states are split by a $\Lambda$-doublet
coupling of 1.667 GHz. In an applied electric field, $\left|e\right\rangle $
and $\left|f\right\rangle $ are not strictly good quantum numbers.
However, they are still useful as labels of a state's Stark character:
$\left|e\right\rangle $ states are strong-electric-field-seeking
while $\left|f\right\rangle $ states are weak-field-seeking.

\begin{figure}[t]
\begin{centering}
\includegraphics[width=1\columnwidth]{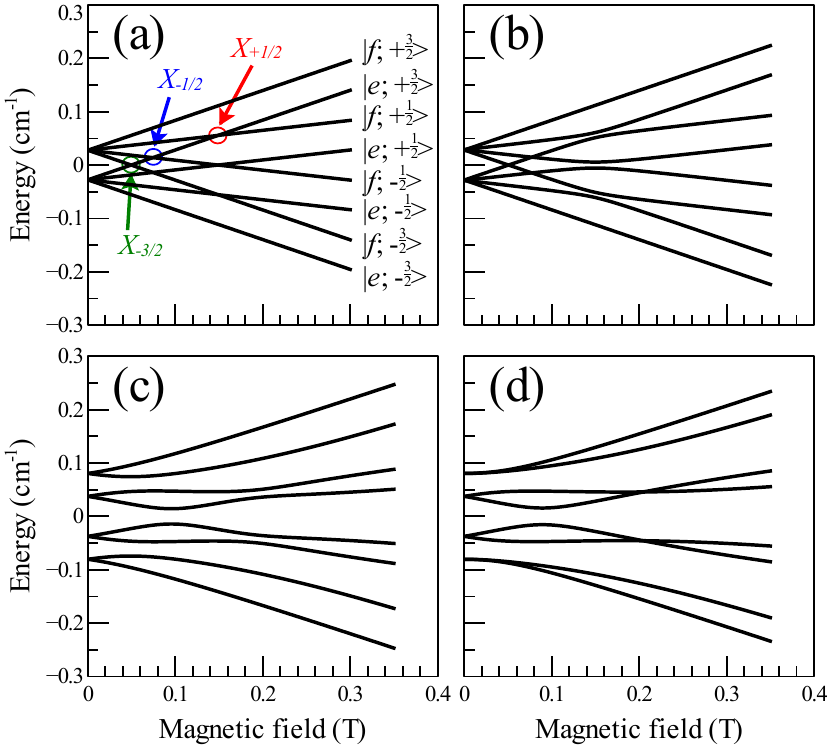}
\par\end{centering}

\caption{\label{fig:Zeeman spectra}(color online) The Zeeman structure of
OH with bias electric fields: (a) 0 V/cm, (b) 500 V/cm at a $90^{\circ}$
magnetic-to-electric field angle $\theta_{EB}$, (c) 5000 V/cm at
$65^{\circ}$, and (d) 5000 V/cm at $90^{\circ}$. The crossings $X_{M}$
of the $\left|e;\,+\frac{3}{2}\right\rangle $ state with the three$\left|f;\, M=\{-\frac{3}{2},\,-\frac{1}{2},\,+\frac{1}{2}\}\right\rangle $
states are labeled in (a).}
\end{figure}

A representative Hamiltonian for $\Lambda$-doublet molecules in mixed
electric and magnetic fields has been derived in \cite{PhysRevA.78.033433}.
While realistic systems must be solved numerically, the basic structure
of the combined $H_{total}=H_{zero-field}+H_{Stark}+H_{Zeeman}$ Hamiltonian
is simple: the zero-field part generates diagonal matrix elements
in a suitable basis (e.~g. Hund's case A fine-structure plus parity
basis $\left|JM\bar{\Omega}\epsilon\right\rangle $ for OH, where
$\epsilon$ = $e$ or $f$ is the $J$-relative parity label), as
does the Zeeman part if one takes the space-fixed $Z$ axis to lie
along the magnetic field $\vec{B}$. The off-diagonal elements are
then purely from the Stark portion. The component of the electric
field $\vec{E}$ parallel to $\vec{B}$ generates matrix elements
which preserve $M$ and change the parity $\epsilon$, while the transverse
components change $M$ by $\pm1$ as well as $\epsilon$. As can be
seen in the Zeeman spectrum of OH (Fig. \ref{fig:Zeeman spectra}(a)),
all states in the lower parity manifold except for the absolute ground
state have Zeeman-induced crossings with one or more of the upper
parity states: in an E-field, these real crossings are coupled by
the off-diagonal Stark elements and become avoided. Working in the
perturbative limit, the crossings $X_{M}$ of Fig. \ref{fig:Zeeman spectra}
have widths $\delta_{+1/2}\propto\left|E\,\sin\theta_{EB}\right|$,
$\delta_{-1/2}\propto\left|E^{3}\,\sin^{2}\theta_{EB}\cos\theta_{EB}\right|$,
and $\delta_{-3/2}\propto\left|E^{3}\,\sin^{3}\theta_{EB}\right|$,
where $\theta_{EB}$ is the angle between $\vec{E}$ and $\vec{B}$.
Since the off-diagonal elements only couple states of opposite parity,
the total perturbation theory order must be odd: crossings with total
$\Delta M$ even such as $X_{-1/2}$ require one extra $\Delta M=0$ coupling
order so that the total coupling changes $\epsilon$ as it must. This
mix of $\Delta M=\pm1$ and $\Delta M=0$
matrix elements in turn leads to a $\left(\sin\theta_{EB}\right)^{\Delta M}\cos\theta_{EB}$ 
angular dependence in the coupling strengths for these even crossings, while the angular
dependence for crossings with $\Delta M$ odd is simply $\left(\sin\theta_{EB}\right)^{\Delta M}$.
Perturbation theory begins to fail, however, at relatively small $E$-fields
(much smaller than the polarizing $E$-field at zero \textbf{$B$}-field)
because of the strength of the electric dipole coupling and the fact
that the Zeeman crossings are gapless at zero $E$-field. Therefore,
for all calculations described in this Letter we diagonalize the full
$8\times8$ ground state fine-structure Hamiltonian of OH \cite{Supplemental,SawyerThesis}.

\begin{figure}[t]
\begin{centering}
\includegraphics[width=1\columnwidth]{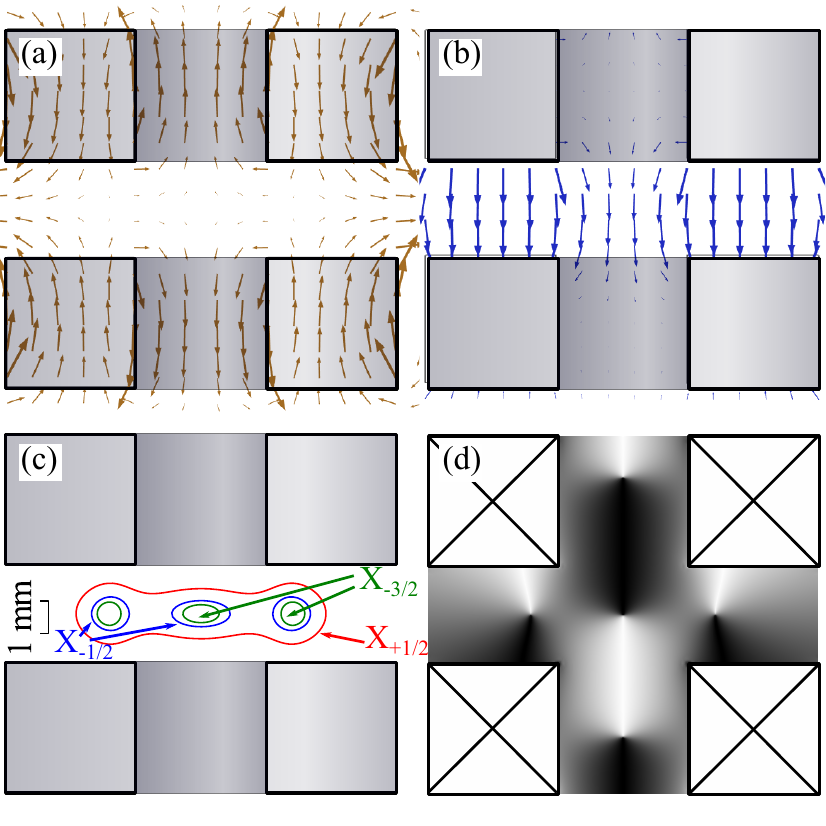}
\par\end{centering}

\caption{\label{fig:trap geometry}(color online, to scale) (a) The magnetic
field distribution in the permanent magnet quadrupole trap. (b). The
electric field distribution created by applying a potential difference
to the trap magnets. (c) The spatial location of the avoided crossings
$X_{M}$. (d) The relative angle $\theta_{EB}$ between electric and
magnetic fields in the trap. White denotes $0^{\circ}$ while black
denotes $180^{\circ}$, implying that the avoided crossings are widest
in the gray regions.}
\end{figure}

We observe these avoided crossings by their dynamical effects on a
sample of magnetically trapped OH molecules. Our Stark deceleration
and magnetic trapping system has been described elsewhere \cite{PhysRevLett.101.203203}.
Briefly, OH molecules are created in an electric discharge through
water vapor in a supersonic expansion and then Stark decelerated and
brought to rest in a permanent magnetic quadrupole trap (Fig. \ref{fig:trap geometry}(a))
at a temperature of $\sim70$ mK, in their $\left|f;\,+\frac{3}{2}\right\rangle $
state. By applying chirped microwave fields to the trap magnet surfaces,
we can then transfer the trapped molecules to the $\left|e;\,+\frac{3}{2}\right\rangle $
state, wherein we can probe the three crossings labeled in Fig. \ref{fig:Zeeman spectra}(a);
application of a DC voltage to the magnets results in a bias E-field
with the distribution shown in Fig. \ref{fig:trap geometry}(b). The
spatial loci of the three crossings in the actual field gradient of
our trap are shown in Fig. \ref{fig:trap geometry}(c). The multiple-well
structure is a result of our use of permanent magnets: the hole in
the center of the magnets allows flux to exit both inside and outside
the magnet ring, yielding a toroidal field minimum between the faces
of the magnet, as well as the central quadrupole trap. The distribution
of $\theta_{EB}$ angles generated by charging the magnets is plotted
in Fig. \ref{fig:trap geometry}(d).

As our trapped molecules are moving at velocities $\lesssim10$ m/s,
only a very small energy gap is needed for each molecule to follow
the adiabats rather than the diabats of the potential surfaces. As
can be seen in Fig. \ref{fig:Zeeman spectra}, the adiabats of all
the levels below the {}``doubly-stretched'' $\left|f;\,+\frac{3}{2}\right\rangle $
state are substantially less trapped than the $\left|e;\,+\frac{3}{2}\right\rangle $
diabat -- or in some cases, are outright antitrapped. Thus, if molecules
are transferred to the $\left|e;\,+\frac{3}{2}\right\rangle $ diabat
in zero field and then a bias field is applied, some fraction of the
molecules will leak out of the trap as they follow the adiabats rather
than Landau-Zener hopping \cite{PhysRevA.23.3107} to follow the trap
diabat. At low fields, the Landau-Zener hopping probability is still
substantial, and so it takes multiple trap oscillations for a molecule
to leak out. At higher fields, however, effectively all molecules
follow the adiabats and so the loss rate is simply limited by the
trap dynamical time $\tau_{dyn}=\frac{1}{4}\nu_{trap}^{-1}$ where
$\nu_{trap}$ is the trap oscillation frequency.

In order to transfer population from $\left|f;\,+\frac{3}{2}\right\rangle $
to $\left|e;\,+\frac{3}{2}\right\rangle $, we use an Adiabatic Rapid
Passage (ARP) technique \cite{Feynman1957} via microwave fields applied
directly to the trap magnets. As the magnets also experience $\pm14$
kV potentials as part of the trap loading sequence, the microwave
system is AC-coupled to the trap by a pair of in-vacuum high voltage
capacitors \cite{Supplemental}. This capacitive coupling stage reduces
the $\pm14$ kV bias to a few volts; two 1.2 GHz high-pass filters
further suppress both the near-DC bias and the switching transients.
By chirping our microwave frequency through the entire range of differential
Zeeman shifts \cite{PhysRevA.74.061402} in our trap over a period
$\tau_{ARP}\ll\tau_{dyn}$, we are able to achieve transfer of $\sim75\,\%$
of the observable trap population. (Since the transfer is a $\pi$
transition, only the fraction of $\vec{E}_{RF}\parallel\vec{B}$ is
useful. Thus, molecules in the gray regions of Fig. \ref{fig:trap geometry}(d)
see much smaller effective intensities than molecules in the white
or black regions.) 

\begin{figure}[t]
\begin{centering}
\includegraphics[width=1\columnwidth]{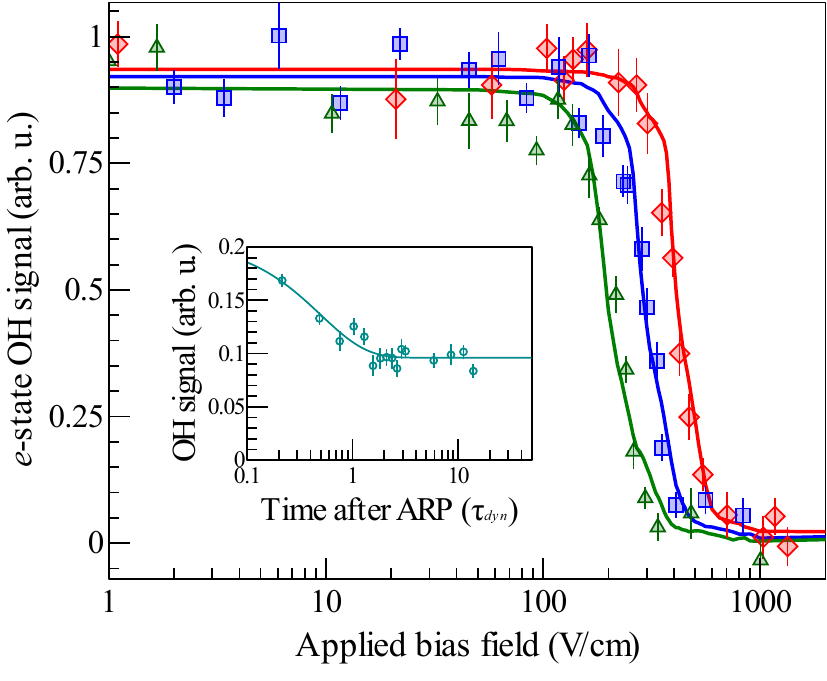}
\par\end{centering}

\caption{\label{fig:ARP}(color online) Post-ARP $e$-state population versus
applied bias field, with 10 (green triangles), 20 (blue squares),
and 40 MHz (red diamond) ARP chirp widths. Solid lines are Monte-Carlo
simulations of the population transfer using realistic bias and RF
field distributions from finite-element calculations, a semiclassical
treatment of the Stark detunings, and a homogeneous stray field as
a free parameter. (inset) Time-of-flight of $e$-state population
immediately after ARP, with no applied bias field. The rapid loss
through the $X_{+1/2}$ crossing is clearly visible: the solid line
is a fit to a single-exponential loss at a rate of $2\tau_{dyn}^{-1}$.}
\end{figure}

In addition to causing dynamical trap loss, the avoided crossings
$X_{M}$ also reduce the efficiency of our ARP pulses: in the vicinity
of the avoided crossings, not only are the states shifted away from
the nominal 1.667 GHz transition frequency, but also the two parity
states are mixed by the electric field. The net result of this mixing
is that molecules transferred near said crossings may not in fact
be transferred to the $\left|e;\,+\frac{3}{2}\right\rangle $ diabat
but rather may end up on the $\left|f;\, M\right\rangle $ diabat
and thus be invisible to our state-selective detection. Figure \ref{fig:ARP}
displays the $\left|e;\,+\frac{3}{2}\right\rangle $ population as
a function of the DC bias field applied during the ARP. The bias field
leads to a rapid drop in transferred population as the $\left|f;\,+\frac{3}{2}\right\rangle \rightarrow\left|e;\,+\frac{3}{2}\right\rangle $
transition is Stark-shifted out of resonance; the bias field required
for this shift increases with the frequency width of the ARP pulse.
We have performed simulations using a numerical solution of the Optical
Bloch Equations to calculate a position dependent ARP transfer probability
with zero electric field. We then calculate the position-dependent
Stark-Zeeman shift by diagonalizing the OH Hamiltonian and convolve
the resulting frequency distribution with the ARP bandwidth to determine
the total fraction transferred. The simulation results shown in Fig.
\ref{fig:ARP} give a good match to the population measurements, showing
that the ARP process is well-understood \cite{Supplemental}. The simulations
suggest the presence of a stray E-field in the trap region on the
order of 100--150 V$/$cm.

\begin{figure}[t]
\begin{centering}
\includegraphics[width=1\columnwidth]{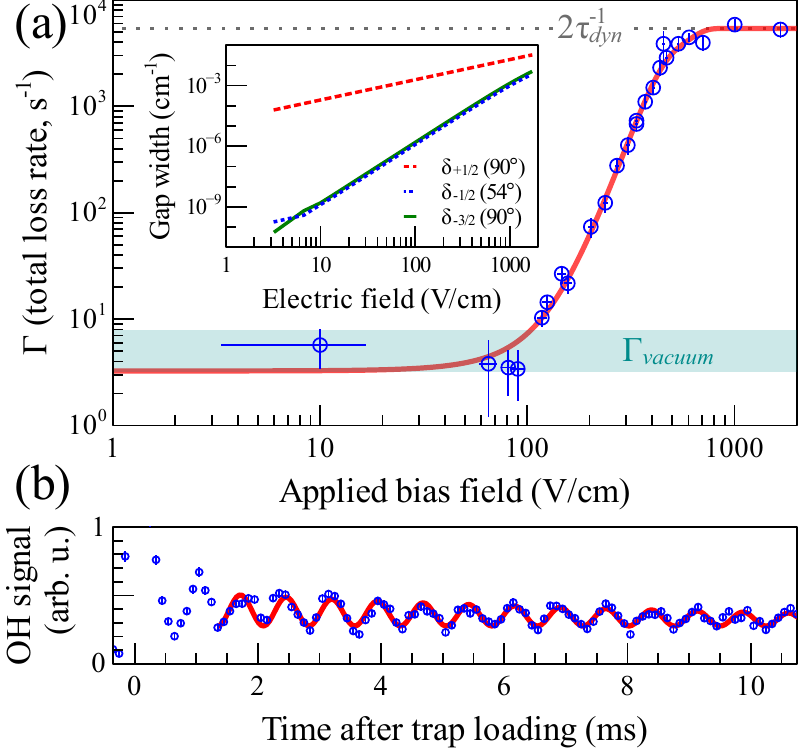}
\par\end{centering}

\caption{\label{Landau-Zener loss}(color online) (a) Loss rate of $\left|e;\, M=+\frac{3}{2}\right\rangle $
molecules versus applied electric field. The solid red line is a fit
to a simple Landau-Zener model; the green shaded stripe denotes the
range of observed loss rates for $\left|f;\,+\frac{3}{2}\right\rangle $
molecules. The bias field is turned on 5 ms after the ARP pulse is
completed, and thus well after the prompt loss seen in the inset of
Fig. \ref{fig:ARP} is completed. (inset) Widths $\delta_{M}$ of
the avoided crossings between $\left|e;\,+\frac{3}{2}\right\rangle $
and $\left|f;\, M\right\rangle $ versus applied electric field, at
their respectively maximizing values of $\theta_{EB}$. $\delta_{+1/2}$
is linear with field, while $\delta_{-1/2}$ and $\delta_{-3/2}$
are cubic. (b) Time-of-flight trace of trap oscillations, observed
by loading the trap with a non-zero center-of-mass velocity. The solid
red fit yields a trap frequency $\nu_{trap}=675$ Hz.}
\end{figure}

Applied E-fields also have a profound effect on trap dynamics. A loss
of $\sim40\%$ of the $e$-state population over $\tau_{dyn}$ is
observed immediately after the ARP pulse (Fig. \ref{fig:ARP} inset),
almost certainly due to prompt loss of molecules following an adiabat
through the $X_{+1/2}$ crossing (which is fully avoided in just the
observed stray E-field) and exiting the trap. At larger bias fields,
loss rates up to $2\tau_{dyn}^{-1}$ are observed with critical fields
again on the order of 200-300 V/cm. The observed loss rates (Fig.
\ref{Landau-Zener loss}(a)) are well-fit by a probabilistic model
\begin{eqnarray*}
\Gamma & = & \Gamma_{vacuum}+\tau_{dyn}^{-1}\:\times\!\!\!\!\!\!\!\!\sum_{{\scriptscriptstyle \PiStrut M\in\{-\frac{3}{2},-\frac{1}{2}\}}}\!\!\!\!\!\!\!\!\!\left(1-\mathcal{P}_{LZ}^{{\scriptscriptstyle (M)}}\right)
\end{eqnarray*}
where $\Gamma_{vacuum}$ is the background vacuum loss rate and 
\begin{eqnarray*}
\mathcal{P}_{LZ}^{{\scriptscriptstyle (M)}} & = & \exp\left(-\frac{2\pi\xi\,\left\langle \delta_{M}^{2}\left(E_{eff}\right)\right\rangle }{\hbar\left(\Delta\mu_{eff}\left(M\right)\right)\,\left\langle \frac{dB}{dt}\right\rangle }\right)
\end{eqnarray*}
is the Landau-Zener probability that the molecule stays on the $\left|e;\,+\frac{3}{2}\right\rangle $
diabat through the crossing $X_{M}$. Angle brackets denote population
averages: $\left\langle \frac{dB}{dt}\right\rangle =\left\langle \vec{\nabla}B\cdot\vec{v}\right\rangle $
is the rate of change of the magnetic field and $\left\langle \delta_{M}\left(E_{eff}\right)\right\rangle =\left\langle \delta_{M}\left(\sqrt{E^{2}+E_{stray}^{2}}\right)\right\rangle $
is the width of the crossing $X_{M}$ as a function of applied electric
field $E$ and stray field $E_{stray}$, while $\Delta\mu_{eff}\left(M\right)=\left(\frac{\nicefrac{3}{2}\cdot\nicefrac{3}{2}}{\nicefrac{3}{2}\cdot\nicefrac{5}{2}}-\frac{\nicefrac{3}{2}\cdot M}{\nicefrac{3}{2}\cdot\nicefrac{5}{2}}\right)\cdot2\mu_{B}$
is the difference between the effective magnetic moments of the $\left|f;\,+\frac{3}{2}\right\rangle $
and $\left|e;\, M\right\rangle $ states. The model contains only
3 free parameters: $\Gamma_{vacuum}$, $E_{stray}$, and $\xi\approx0.12$,
which incorporates the errors in our estimates of the population-averaged
parameters. (While we can estimate $\left\langle \frac{dB}{dt}\right\rangle $
reasonably well from Monte-Carlo trap dynamics simulations, $\left\langle \delta_{M}\left(E_{eff}\right)\right\rangle $
has a very fast variation with $\left\langle \theta_{EB}\right\rangle $
and so is hard to accurately calculate. It is therefore responsible
for most of $\xi$'s deviation from unity.) $\Gamma_{vacuum}$ can
be independently observed from the $f$-state lifetime, while $E_{stray}$
can be derived from Fig. \ref{fig:ARP}; both of these independent
measurements are consistent with the fit. $\tau_{dyn}$ is directly
measurable from the trap oscillation frequency $\nu_{trap}$ observed
in Fig. \ref{Landau-Zener loss}(b). As both $\delta_{-1/2}$ and
$\delta_{-3/2}$ are third-order in $E$, the most striking feature
of the Landau-Zener model is that it explains the extremely steep
rise in $\Gamma$: $\Gamma\sim1-\exp\left(-\alpha E^{6}\right)$ where
$\alpha$ represents the Landau-Zener parameters.

The success of this model prompts two conclusions. The first is that
magnetic trapping of the lower parity doublet in any molecule is fraught
with difficulties: in OH, even just the fields from stray charge accumulating
on the trap mounts is enough to cause unity loss through the $X_{+1/2}$
channel. (We are currently implementing a new trap mount design to
minimize the patch charges in our future work.) In general, if one
wishes to explore polar radicals in mixed fields in any states except
the doubly-stretched ones, either optical trapping or an Ioffe-Pritchard
trap with a sufficiently large offset $B$-field as to be above all
the crossings is necessary; this implies that polar radicals in their
lower parity doublet are extremely challenging candidates for slowing
with a magnetic coilgun \cite{PhysRevA.75.031402,PhysRevA.77.051401}.
More broadly, in precision measurement applications it is imperative
to take a close look \cite{PhysRevA.80.023418} at the exact Stark-Zeeman
spectrum of the system of interest, as the spectral distortions from
the avoided crossings can persist over very large ranges in magnetic
field.
\begin{acknowledgments}
This work was funded by DOE, the AFOSR MURI on Cold Molecules, NSF,
and NIST. M. Hummon is an National Research Council postdoctoral fellow.
\end{acknowledgments}

\bibliographystyle{apsrev4-1}
%

\end{document}


\title{Supplemental Material to {}``Microwave state transfer and adiabatic
dynamics of magnetically trapped polar molecules''}

\author{Benjamin K. Stuhl}
\email{stuhl@jila.colorado.edu}

\author{Mark Yeo}

\author{Brian C. Sawyer}
\altaffiliation{Present address: Ion Storage Group, National Institute of Standards and Technology, Boulder, CO 80305, USA}

\author{Matt Hummon}
\author{Jun Ye}

\affiliation{JILA, National Institute of Standards and Technology and University
of Colorado, Department of Physics, 440 UCB, Boulder, Colorado 80309,
USA}

\maketitle

\subsection{Derivation of the Stark-Zeeman Hamiltonian}

\global\long\def\threej#1#2#3#4#5#6{\left(\begin{array}{ccc}
 #1  &  #2  &  #3\\
#4  &  #5  &  #6 
\end{array}\right)}
In order to evaluate molecule-field interaction matrix elements of
the form $H_{i}=-\vec{\mu}\cdot\vec{\mathcal{F}}$ (where $\vec{\mathcal{F}}$
represents either $\vec{B}$ or $\vec{E}$), it is necessary to transform
between the molecule-fixed frame in which the appropriate dipole moment
$\vec{\mu}$ is defined and the laboratory frame in which $\vec{\mathcal{F}}$
is defined. Following Lara et al. \cite{PhysRevA.78.033433}, we use
the Wigner rotation matrix elements $\mathcal{D}_{qk}^{1*}\left(\omega\right)$
to write
\[
H_{i}=-\vec{\mu}\cdot\vec{\mathcal{F}}=-\sum_{q}\left(-1\right)^{q}\left(\sum_{k}\mathcal{D}_{qk}^{1*}\left(\omega\right)\mu_{k}\right)\mathcal{F}_{-q}.
\]
While OH is not fully a Hund's case (a) molecule, it is nonetheless
close enough that both the electric and magnetic dipole moments may
be considered to lie along the internuclear axis and so $H_{i}$ simplifies
to 
\[
H_{i}=-\sum_{q}\left(-1\right)^{q}\mathcal{D}_{q0}^{1*}\left(\omega\right)\mu_{k=0}\mathcal{F}_{-q}.
\]
If $\vec{B}$ and $\vec{E}$ are not parallel, it is further necessary
to rotate the laboratory fields to a combined frame. Without loss
of generality, we choose Z along $\vec{B}$ and transform $\vec{E}$
using the reduced rotation matrix elements $d_{qk}^{1*}\left(\theta_{EB}\right)$
to write 

\[
H_{i}=-\mathcal{D}_{00}^{1*}\left(\omega\right)\mu_{k=0}^{(B)}\left|B\right|-\sum_{q}\mathcal{D}_{q0}^{1*}\left(\omega\right)\mu_{k=0}^{(E)}d_{-q0}^{1*}\left(\theta_{EB}\right)\left|E\right|.
\]
Using Lara et al.'s results for the $\left|JM\bar{\Omega}\epsilon\right\rangle $
parity basis yields the matrix elements
\begin{eqnarray*}
\left\langle JM\bar{\Omega}\epsilon\right|H_{i}\left|J'M'\bar{\Omega}'\epsilon'\right\rangle  & = & \sqrt{2J+1}\sqrt{2J'+1}\threej J1{J'}{-\bar{\Omega}}0{\bar{\Omega}'}\left(-1\right)^{M-\bar{\Omega}}\\
 &  & \times\left\{ \left(\frac{1+\epsilon\epsilon'\left(-1\right)^{J+J'+2\bar{\Omega}}}{2}\right)\threej J1{J'}{-M}0{M'}\mu_{B}B\left(\Lambda+g_{e}\Sigma\right)\right.\\
 &  & +\left.\left(\frac{1-\epsilon\epsilon'\left(-1\right)^{J+J'+2\bar{\Omega}}}{2}\right)\sum_{q}\threej J1{J'}{-M}q{M'}d_{-q0}^{1*}\left(\theta_{EB}\right)\mu_{e}E\right\} 
\end{eqnarray*}
where $g_{e}$ is the electron Land\'e $g$-factor and $\mu_{e}$
is the molecular electric dipole moment.

\subsection{Full OH Hamiltonian}

Evaluating the matrix elements described previously for the OH $\left|J=\frac{3}{2},\, M,\,\bar{\Omega}=\frac{3}{2},\,\epsilon\right\rangle $
ground state manifold yields the $8\times8$ matrix used for combined
Stark-Zeeman calculations in this paper:\global\long\def\dd{\frac{\Delta}{2}}
\global\long\def\ubb{\mu_{B}B}
\global\long\def\uec{\mu_{e}E\,\cos\theta_{EB}}
\global\long\def\ues{\mu_{e}E\,\sin\theta_{EB}}
{\tiny 
\[
\left(\begin{array}{cccccccc}
-\dd-\frac{6}{5}\ubb & 0 & 0 & 0 & \frac{3}{5}\uec & -\frac{\sqrt{3}}{5}\ues & 0 & 0\\
0 & -\dd-\frac{2}{5}\ubb & 0 & 0 & -\frac{\sqrt{3}}{5}\ues & \frac{1}{5}\uec & -\frac{2}{5}\ues & 0\\
0 & 0 & -\dd+\frac{2}{5}\ubb & 0 & 0 & -\frac{2}{5}\ues & -\frac{1}{5}\uec & -\frac{\sqrt{3}}{5}\ues\\
0 & 0 & 0 & -\dd+\frac{6}{5}\ubb & 0 & 0 & -\frac{\sqrt{3}}{5}\ues & -\frac{3}{5}\uec\\
\frac{3}{5}\uec & -\frac{\sqrt{3}}{5}\ues & 0 & 0 & \dd-\frac{6}{5}\ubb & 0 & 0 & 0\\
-\frac{\sqrt{3}}{5}\ues & \frac{1}{5}\uec & -\frac{2}{5}\ues & 0 & 0 & \dd-\frac{2}{5}\ubb & 0 & 0\\
0 & -\frac{2}{5}\ues & -\frac{1}{5}\uec & -\frac{\sqrt{3}}{5}\ues & 0 & 0 & \dd+\frac{2}{5}\ubb & 0\\
0 & 0 & -\frac{\sqrt{3}}{5}\ues & -\frac{3}{5}\uec & 0 & 0 & 0 & \dd+\frac{6}{5}\ubb
\end{array}\right).
\]
}{\tiny \par}

\subsection{Microwave coupling circuit design}

\begin{center}
\begin{figure}[H]
\begin{centering}
\includegraphics{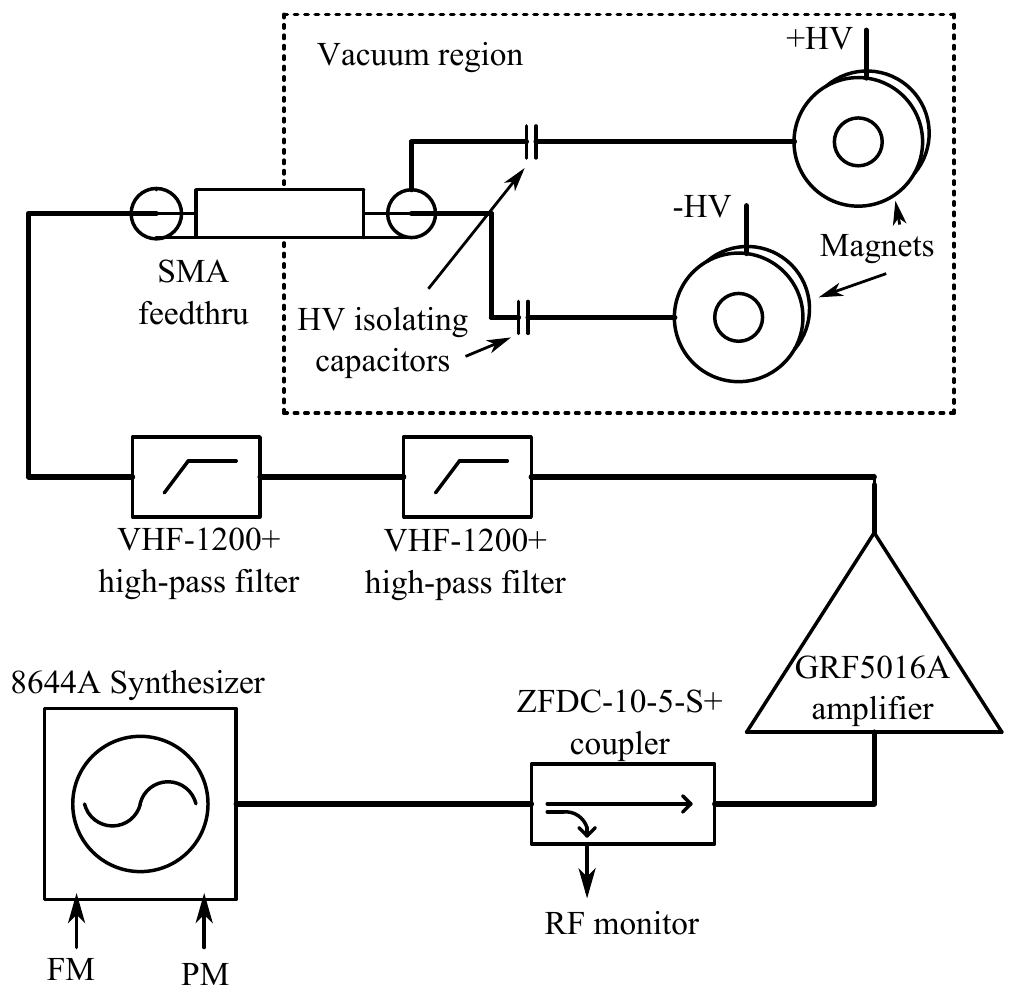}
\par\end{centering}

\caption{\label{fig:microwaves}A schematic of the microwave system use to
transfer OH between $\Lambda$-doublet levels.}
\end{figure}

\par\end{center}

In order to couple microwave power on to our trap magnets while protecting
the microwave system from the $\pm14$ kV applied to the magnets during
the trap loading sequence, we use a pair of custom in-vacuum capacitors
to block the DC high voltage plus two more high-pass filters (Mini-Circuits
VHF-1200+%
\footnote{Manufacturer and part names are listed for completeness only and do
not constitute an endorsement by NIST.%
}) to block the AC transients generated by the magnet switching.

Tunable microwave power is generated by an HP 8644A synthesizer, which
is programmed to form the ARP pulse by voltages supplied to its frequency
and pulse modulation inputs. A Mini-Circuits ZFDC-10-5-S+ directional
coupler allows monitoring of the microwave signal before it is amplified
by a GTC GRF5016A power amplifier. The microwaves are then coupled
through the protection filters and into the vacuum system via an SMA
feedthru. Typical power levels are 30.4--35.4 dBm (1--3 W) at the
feedthru. As the ARP chirp time $\tau_{ARP}$ must be much shorter
than $\tau_{dyn}$, $\tau_{ARP}$ is generally between 20 and 80 $\mathrm{\mu s}$.

\subsection{Numerical simulations of ARP}

\begin{figure}[H]
\begin{centering}
\includegraphics{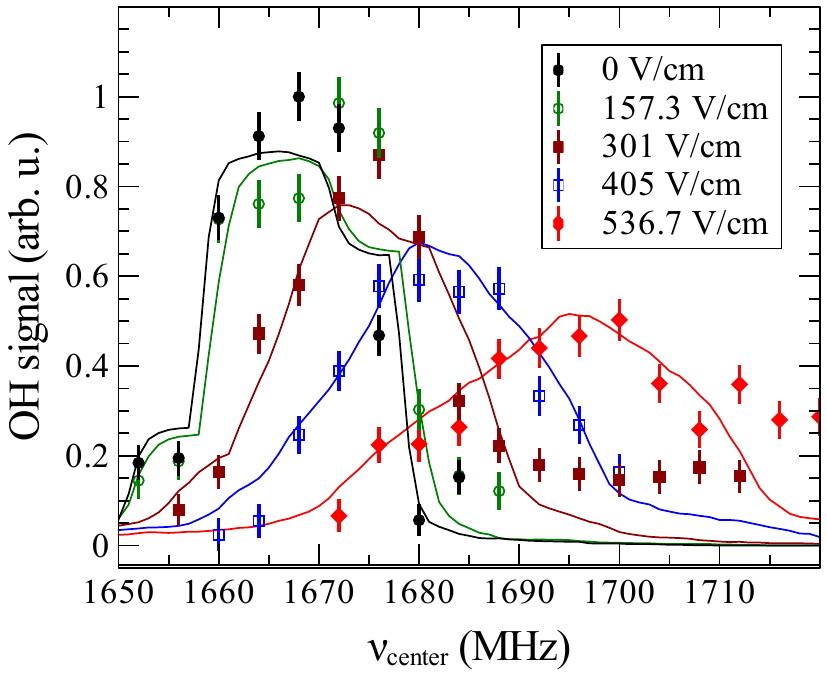}
\par\end{centering}

\caption{\label{fig:ARP spectra}ARP spectra at several different bias $E$-fields.
Points are data while solid lines are simulation results. The Stark-Zeeman
broadening due to the avoided crossings is clearly visible.}
\end{figure}

As described in the main text, we have performed numerical simulations
of the ARP population dynamics. The simulation code works as follows:
\begin{enumerate}
\item For each molecule in a Monte Carlo distribution, the Optical Bloch
Equations are solved using the local magnitude of the microwave field
(the microwave field is assumed to follow the DC field distribution
generated by charging the magnet surfaces) and its angle against the
local magnetic field. Both field distributions were calculated using
a finite-element package. The final probability of the molecule transferring
to the lower $\Lambda$-doublet state is then recorded.
\item For each molecule, the local static $E$-field is calculated and then
combined with the local $B$-field to find the total Stark-Zeeman
detuning. This yields a transfer probability density as a function
of microwave frequency.
\item The probability density function is then integrated over the range
of frequencies the ARP chirp sweeps through to get a final transfer
probability.
\end{enumerate}
Figure \ref{fig:ARP spectra} shows several ARP spectra at different
bias $E$-fields. The agreement between data (points) and simulation
(solid lines) is good; in particular, the simulations capture the
Stark-Zeeman broadening visible at larger fields. In a homogeneous
field and the absence of avoided crossings, the application of an
$E$-field would merely push the two $\Lambda$-doublet states apart
without broadening the transition. The field in the trap region is
not fully homogeneous, which contributes some broadening, while the
curvature of the $\left|e;\, M=+\frac{3}{2}\right\rangle $ state
near its avoided crossings further increases the Stark-Zeeman broadening.
In order to fit the simulations to the data, a background stray $E$-field
on the order of 120 $\mathrm{V/cm}$ was required, consistent with
the field required for the Landau-Zener model.

\bibliographystyle{plain}